\documentclass{article}

\PassOptionsToPackage{numbers, compress}{natbib}
\usepackage[preprint]{neurips2023}
\usepackage[utf8]{inputenc} % allow utf-8 input
\usepackage[T1]{fontenc}    % use 8-bit T1 fonts
\usepackage{hyperref}       % hyperlinks
\usepackage{url}            % simple URL typesetting
\usepackage{booktabs}       % professional-quality tables
\usepackage{amsfonts}       % blackboard math symbols
\usepackage{nicefrac}       % compact symbols for 1/2, etc.
\usepackage{microtype}      % microtypography
\usepackage{xcolor}         % colors
\usepackage{natbib} 
\usepackage{graphicx}
\usepackage{amsmath}

\title{Music- and Lyrics-driven Dance Synthesis}
\author{%
    Wenjie Yin\textsuperscript{1}, % yinw@kth.se 
    Qingyuan Yao\textsuperscript{2}, % qingyuan.yao@etu.sorbonne-universite.fr
    Yi Yu\textsuperscript{3}\thanks{The corresponding author}\space, % yiyu@nii.ac.jp
    Hang Yin\textsuperscript{4},  % hayi@di.ku.dk
    Danica Kragic\textsuperscript{1},  % dani@kth.se
    Mårten Björkman\textsuperscript{1}\\   % celle@kth.se
    KTH Royal Institute of Technology\textsuperscript{1}, 
    Sorbonne University\textsuperscript{2},\\
    National Institute of Informatics\textsuperscript{3}, 
    University of Copenhagen\textsuperscript{4}\\
  \texttt{yinw@kth.se, yiyu@nii.ac.jp} \\
}

\begin{document}

\maketitle

% === ABSTRACT ===
\begin{abstract}
Lyrics often convey information about the songs that are beyond the auditory dimension, enriching the semantic meaning of movements and musical themes. Such insights are important in the dance choreography domain. However, most existing dance synthesis methods mainly focus on music-to-dance generation, without considering the semantic information. To complement it, we introduce JustLMD, a new multimodal dataset of 3D dance motion with music and lyrics. To the best of our knowledge, this is the first dataset with triplet information including dance motion, music, and lyrics. 
Additionally, we showcase a cross-modal diffusion-based network designed to generate 3D dance motion conditioned on music and lyrics. The proposed JustLMD dataset encompasses 4.6 hours of 3D dance motion in 1867 sequences, accompanied by musical tracks and their corresponding English lyrics. 
\end{abstract}

% === INTRODUCTION ===
\section{Introduction}
% What has been done? % Why care?
Recent breakthroughs in generative models, notably in normalizing flows~\cite{papamakarios2021normalizing} and diffusion models~\cite{ho2020denoising}, have significantly advanced applications like music-conditioned dance generation. Such advancements not only enrich the artistic dimension of choreography but also provide valuable insights for dance research~\cite{valle2021transflower, alexanderson2023listen, li2021ai, yin2023multimodal}. 
Given the rising popularity of dance content on digital platforms such as YouTube and TikTok, these pioneering technologies hold vast potential to be integrated into creative processes within the dance domain.

% What is missing?
However, many existing technologies primarily focus on the relationship between music and dance, without considering the integral role of lyrics in dance choreography. While music-conditioned models can already produce realistic rhythmically-aligned dance movements, lyrics offer additional information that can enrich and enhance the semantic meaning of the dance. For instance, there exists a strong linkage between dance motion and song lyrics in modern dance~\cite{powell2019modern}. 
% How do you address this? 
Further exploration is needed to explore integrating both lyrics and music in dance synthesis. To fill this research void, we introduce a multimodal dataset with synchronized dance motion, music, and lyrics. Moreover, we present a cross-modal diffusion model to facilitate dance motion synthesis based on both lyrics and music.

% === METHODS ===
\section{Methods}

In this section, we introduce the preparation pipeline of our dataset and the proposed baseline model for dance synthesis. 

% data
\subsection{Data Preparation Pipeline}

Due to the lack of datasets with dance motion, music, and lyrics information, we created the proposed JustLMD dataset using existing \textit{Just Dance} videos. Ubisoft's  \textit{Just Dance} is a motion-based rhythm dancing game, with its annual releases, and has been a classic in video games. Just Dance engages players by having them mimic the moves of an on-screen dancer.
To prepare our multimodal dataset, we adopted the following pipeline. 

\begin{itemize}
    \item \textbf{Video Conversion}: We transformed YouTube videos of \textit{Just Dance} into .mp4 format.
    \item \textbf{Video Preprocessing}: Utilizing \textit{EasyMocap}~\cite{dong2020motion}, we achieved high-fidelity body estimation from these videos at a rate of 60 fps.
    \item \textbf{Music Extraction}: The music from the videos was saved in .wav format.
    \item \textbf{Lyrics Synchronization}: We manually sourced the lyrics corresponding to each music song and aligned them with the musical timeline.
    \item \textbf{Feature Preparation}:
    \begin{itemize}
        \item \textbf{Pose Representation}: We represent dance as sequences of poses using the 24-joint SMPL format~\cite{loper2015smpl}, using a 6-DOF rotation representation and a 4-dimensional binary foot contact label, resulting in a 151-dimensional feature.
        \item \textbf{Music Feature Extraction}: We employed \textit{librosa}~\cite{mcfee2015librosa} or \textit{Jukebox}~\cite{dhariwal2020jukebox} to extract music features, yielding a 35- or 4800-dimensional feature.
        \item \textbf{Lyrics Feature Embedding}: Lyrics were then processed and embedded into a pre-trained CLIP latent~\cite{radford2021learning} or Bert embedding~\cite{devlin2018bert}, resulting in a 512- or 768-dimensional feature.
    \end{itemize}
\end{itemize}

Our data collection and feature preparation code can be accessed \href{https://github.com/yy1lab/LMD}{here}\footnotemark[1]. Validity of the task and examples of the correlation between motion and lyrics are illustrated in Appendix~\ref{ap:b}. 

% model
\subsection{Multimodal Diffusion Models}
\begin{figure}[ht]
\centering
  \includegraphics[width=0.6\linewidth]{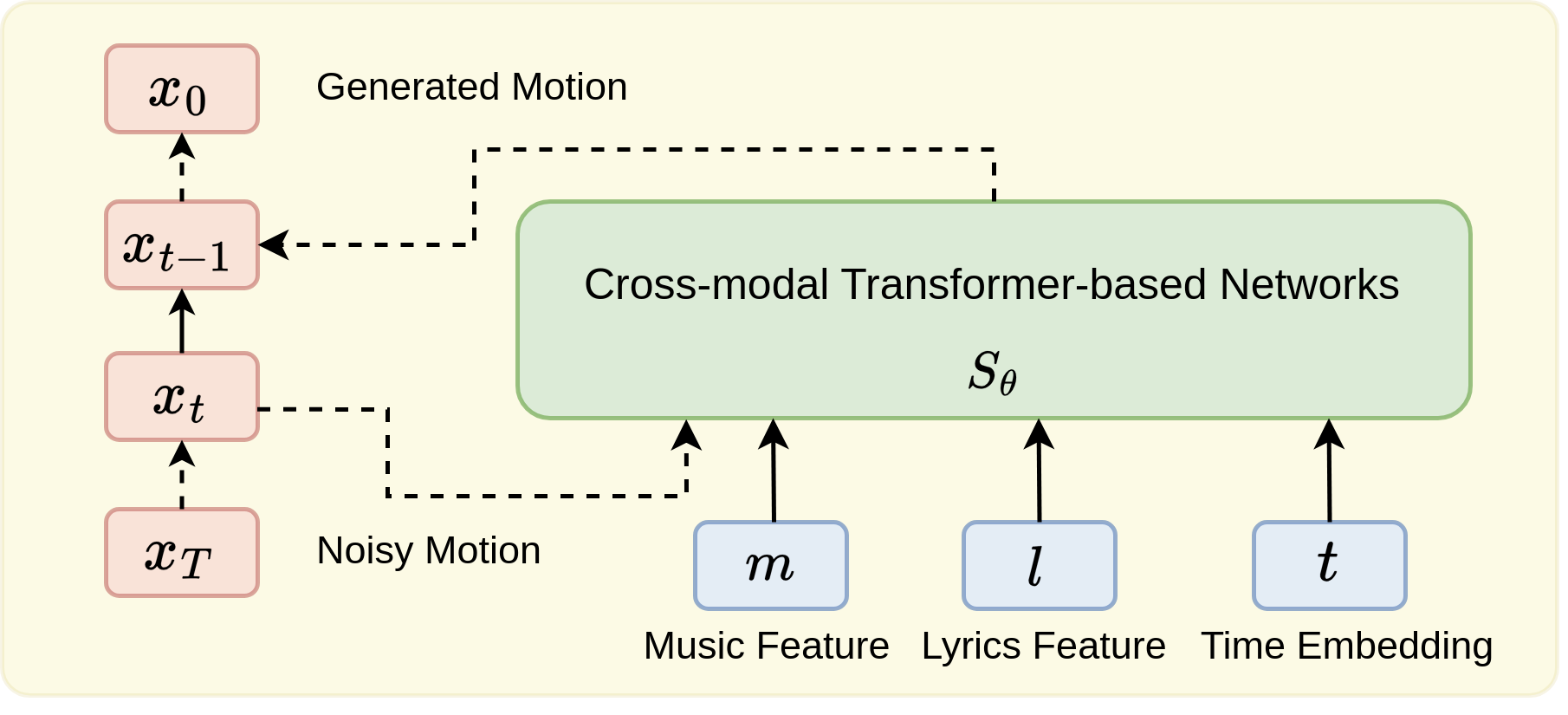}
  \caption{Architecture overview of the cross-modal transformer-based diffusion models, the music features and text features are conditional information that act as cross-attention context. The diffusion model takes noisy sequences and produces the estimated motion sequences. }
  \label{fig:model}
\end{figure}

Our dance motion synthesis framework is built on cross-modal diffusion models. This architecture utilizes a transformer-based diffusion model that accepts conditional feature vectors as input. In this setting, the conditional feature vectors include the music feature and lyrics feature.  It then generates corresponding motion sequences, without autoregression or recurrent connections, as depicted in Figure \ref{fig:model}. The model incorporates a cross-attention mechanism, following~\cite{saharia2022photorealistic}. 
We optimize the $\theta$-parameterized score networks $\boldsymbol{s}_\theta$ with dance motion $\boldsymbol{x}$ paired conditional music feaure $\boldsymbol{m}$ and lyric feaure $\boldsymbol{l}$, $t$ is the time embedding. The objective function is simplified as:
\begin{equation}
    \mathbb{E}_{\boldsymbol{x}, {t}}\left\|\boldsymbol{x} - \boldsymbol{s}_{\theta}(\boldsymbol{x}_{{t}}, {t}, \boldsymbol{m}, \boldsymbol{l}) \right\|^2_2.
\end{equation}

\footnotetext[1]{\url{https://github.com/yy1lab/LMD}}

\section{Discussion}
In this paper, we introduce a framework for dance motion synthesis driven by both music and lyrics, accompanied by a novel dataset collected to bridge the existing research gap. Moving forward, we aim to explore deeper into the influence of the lyrics modality. Building on our present efforts, we anticipate enlarging the dataset for training generative models.

\section{Ethical Implications}
We present a multimodal dataset and dance synthesis method. A primary concern comes from the data source. Our model relies on dance motion extracted from public game videos. These movements, though available publicly, the original creators of the dance motion can claim their copyrights. Furthermore, automating the choreographic process based on existing dances raises questions about originality and creativity in art.

\bibliographystyle{unsrt}
\bibliography{neurips2023}

\begin{thebibliography}{10}

\bibitem{papamakarios2021normalizing}
George Papamakarios, Eric Nalisnick, Danilo~Jimenez Rezende, Shakir Mohamed, and Balaji Lakshminarayanan.
\newblock Normalizing flows for probabilistic modeling and inference.
\newblock {\em The Journal of Machine Learning Research}, 22(1):2617--2680, 2021.

\bibitem{ho2020denoising}
Jonathan Ho, Ajay Jain, and Pieter Abbeel.
\newblock Denoising diffusion probabilistic models.
\newblock {\em Advances in neural information processing systems}, 33:6840--6851, 2020.

\bibitem{valle2021transflower}
Guillermo Valle-P{\'e}rez, Gustav~Eje Henter, Jonas Beskow, Andre Holzapfel, Pierre-Yves Oudeyer, and Simon Alexanderson.
\newblock Transflower: probabilistic autoregressive dance generation with multimodal attention.
\newblock {\em ACM Transactions on Graphics (TOG)}, 40(6):1--14, 2021.

\bibitem{alexanderson2023listen}
Simon Alexanderson, Rajmund Nagy, Jonas Beskow, and Gustav~Eje Henter.
\newblock Listen, denoise, action! audio-driven motion synthesis with diffusion models.
\newblock {\em ACM Transactions on Graphics (TOG)}, 42(4):1--20, 2023.

\bibitem{li2021ai}
Ruilong Li, Shan Yang, David~A Ross, and Angjoo Kanazawa.
\newblock Ai choreographer: Music conditioned 3d dance generation with aist++.
\newblock In {\em Proceedings of the IEEE/CVF International Conference on Computer Vision}, pages 13401--13412, 2021.

\bibitem{yin2023multimodal}
Wenjie Yin, Hang Yin, Kim Baraka, Danica Kragic, and M{\aa}rten Bj{\"o}rkman.
\newblock Multimodal dance style transfer.
\newblock {\em Machine Vision and Applications}, 34(4):1--14, 2023.

\bibitem{powell2019modern}
Hayley~Elizabeth Powell.
\newblock Modern dance choreography: Beyond the movement an analysis between lyrics and movement: Can identities be developed through modern dance choreography?
\newblock {\em Annual Review of Education, Communication \& Language Sciences}, 16(2), 2019.

\bibitem{dong2020motion}
Junting Dong, Qing Shuai, Yuanqing Zhang, Xian Liu, Xiaowei Zhou, and Hujun Bao.
\newblock Motion capture from internet videos.
\newblock In {\em Computer Vision--ECCV 2020: 16th European Conference, Glasgow, UK, August 23--28, 2020, Proceedings, Part II 16}, pages 210--227. Springer, 2020.

\bibitem{loper2015smpl}
Matthew Loper, Naureen Mahmood, Javier Romero, Gerard Pons-Moll, and Michael~J Black.
\newblock Smpl: A skinned multi-person linear model.
\newblock {\em ACM transactions on graphics (TOG)}, 34(6):1--16, 2015.

\bibitem{mcfee2015librosa}
Brian McFee, Colin Raffel, Dawen Liang, Daniel~P Ellis, Matt McVicar, Eric Battenberg, and Oriol Nieto.
\newblock librosa: Audio and music signal analysis in python.
\newblock In {\em Proceedings of the 14th python in science conference}, volume~8, pages 18--25, 2015.

\bibitem{dhariwal2020jukebox}
Prafulla Dhariwal, Heewoo Jun, Christine Payne, Jong~Wook Kim, Alec Radford, and Ilya Sutskever.
\newblock Jukebox: A generative model for music.
\newblock {\em arXiv preprint arXiv:2005.00341}, 2020.

\bibitem{radford2021learning}
Alec Radford, Jong~Wook Kim, Chris Hallacy, Aditya Ramesh, Gabriel Goh, Sandhini Agarwal, Girish Sastry, Amanda Askell, Pamela Mishkin, Jack Clark, et~al.
\newblock Learning transferable visual models from natural language supervision.
\newblock In {\em International conference on machine learning}, pages 8748--8763. PMLR, 2021.

\bibitem{devlin2018bert}
Jacob Devlin, Ming-Wei Chang, Kenton Lee, and Kristina Toutanova.
\newblock Bert: Pre-training of deep bidirectional transformers for language understanding.
\newblock {\em arXiv preprint arXiv:1810.04805}, 2018.

\bibitem{saharia2022photorealistic}
Chitwan Saharia, William Chan, Saurabh Saxena, Lala Li, Jay Whang, Emily~L Denton, Kamyar Ghasemipour, Raphael Gontijo~Lopes, Burcu Karagol~Ayan, Tim Salimans, et~al.
\newblock Photorealistic text-to-image diffusion models with deep language understanding.
\newblock {\em Advances in Neural Information Processing Systems}, 35:36479--36494, 2022.

\end{thebibliography}

\appendix

\section{Validity of the Task}
\label{ap:b}

The task of synthesizing dance from both music and lyrics could be supported by the following considerations:  
\begin{itemize}
    \item \textbf{Semantic Motion Correspondence}:  
    While this argument is intuitive, it doesn't always hold, given that choreography is also influenced by musical composition and the chosen dance style. For instance, the word ``NO'' naturally correlates with the gesture of shaking one's head.   
    \item \textbf{Emotional and Contextual Influence}: 
    Lyrics can reinforce the emotion conveyed by the music. A phrase like ``Break my heart'' might be expressed through a more quiet and sad movement.    
    \item \textbf{Rhythmic Pattern Influence}: For example, repetitive lyrical patterns such as ``Oh oh oh'' or ``Ma-ma-ma'' can suggest a series of recurring dance motions.
    \item \textbf{Distinguishing or Linking Music and Lyrics}: It's not uncommon for different song segments to have the same music but distinct lyrics. Incorporating lyrics in the choreography could prevent generating repetitive motions for similar musical segments from dance synthesis models.
\end{itemize}

\begin{figure}[ht]
\centering
  \centering
    \includegraphics[width=0.8\textwidth]{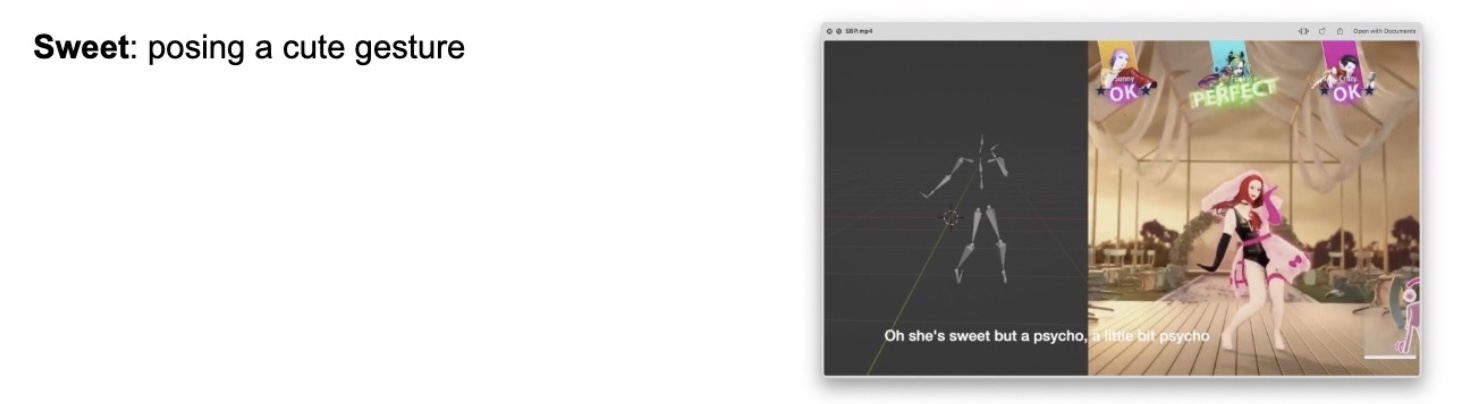}
    \includegraphics[width=0.8\textwidth]{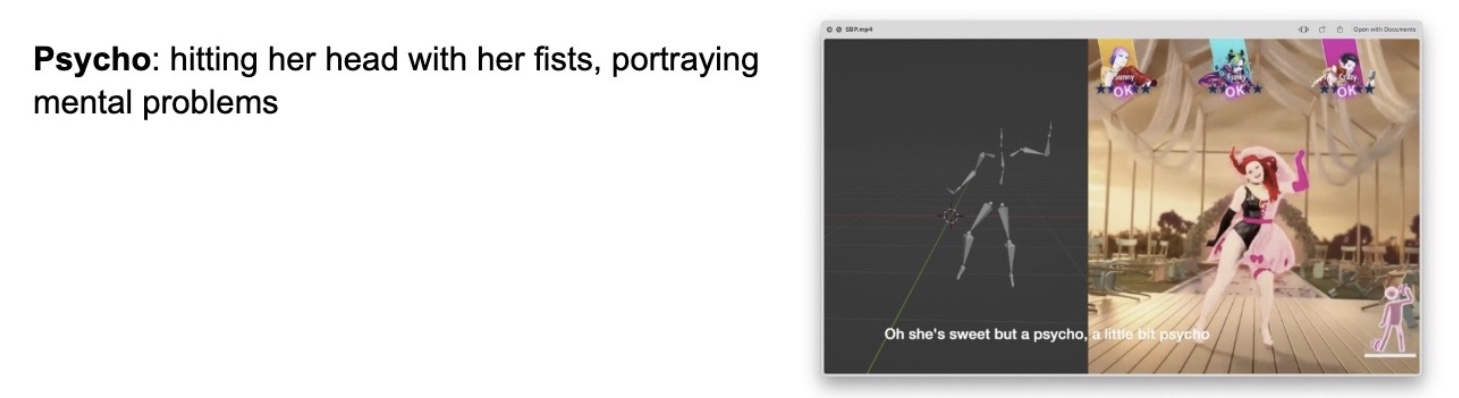}
    \includegraphics[width=0.8\textwidth]{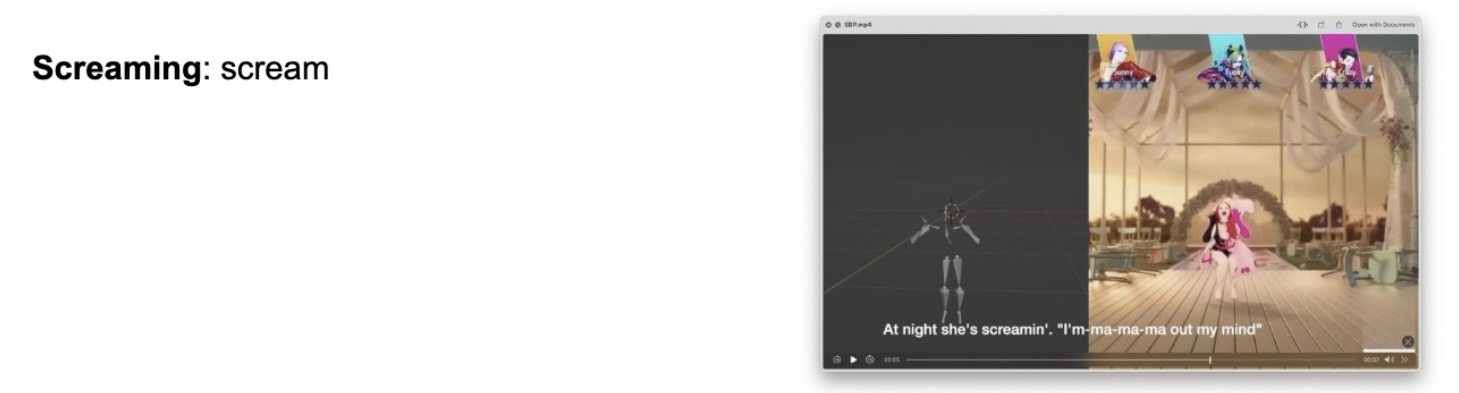}
    \includegraphics[width=0.8\textwidth]{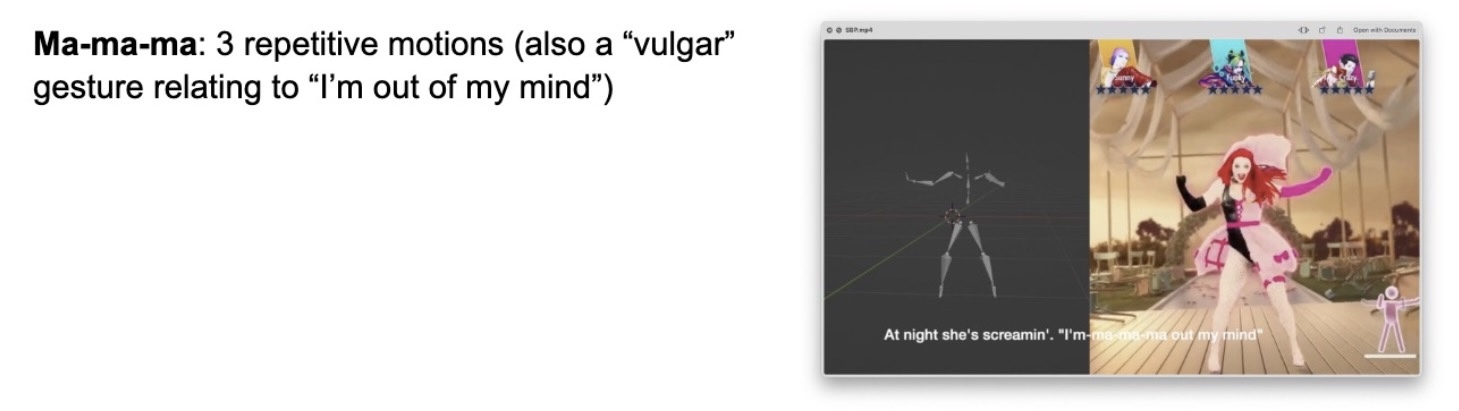}
  \caption{Examples illustrating the connection between lyrics and dance motions, where dancers' movements align closely with lyrical content. }
  \label{fig:lyric-motion}
\end{figure}

In Figure \ref{fig:lyric-motion}, we present examples that demonstrate the correlation between lyrics and corresponding dance motions. Notably, the lyric ``Sweet'' is associated with a charming and cute gesture, while ``screaming'' is visually represented by a shouting pose.

\end{document}